\documentclass{iopart}
\usepackage{iopams}
\usepackage[colorlinks=true,citecolor=blue]{hyperref}
\expandafter\let\csname equation*\endcsname\relax
  \expandafter\let\csname endequation*\endcsname\relax
\usepackage{amsmath,amsthm,amssymb,mathpazo,times,graphicx}
\usepackage[varg]{txfonts}
\usepackage{color}
\makeatletter 
0
\long\def\@makefntext#1{\parindent 1em\noindent
 \makebox[1em][l]{\footnotesize\rm$^\arabic{footnote}$}%
 \footnotesize\rm #1}
\def\@makefnmark{\hbox{$^\arabic{footnote}$}}
\def\@thefnmark{\arabic{footnote}}
\makeatother
\eqnobysec
\allowdisplaybreaks

\let\^=\hat
\let\=\bar
\def\@#1{\mathbf{#1}}
\def\bfnabla{{\mathbb\nabla}}

\def\const{\mathrm{const}}
\def\cn{\mathop{\rm cn}\nolimits}

\def\K{K_m}
\def\E{E_m}
\newcommand\partialderiv[3][]{\frac{\partial^{#1}#2}{\partial {#3}^{#1}}}
\newcommand\deriv[3][]{\frac{d^{#1}#2}{d{#3}^{#1}}}
\def\DDy#1{\frac{D#1}{Dy}}
\def\txtfrac#1#2{{\textstyle\frac{#1}{#2}}}
\def\diag{\mathop{\rm diag}\nolimits}
\let\~=\tilde

\def\be{\begin{equation}}
\def\ee{\end{equation}}
\def\bea{\begin{eqnarray}}
\def\eea{\end{eqnarray}}
\def\[{\begin{eqnarray}}
\def\]{\end{eqnarray}}
\def\bse{\begin{subequations}}
\def\ese{\end{subequations}}

\advance\textwidth 4mm
\advance\hoffset -2mm
\advance\textheight 4mm
\advance\voffset -2mm

\advance\parskip 1pt

\def\bb{\relax}
\def\eb{\relax}

\def\em{\endgroup}

\def\u{\=u}
\def\v{\=v}
\def\k{k}

\begin{document}

\title{Integrability, exact reductions and special solutions of the KP-Whitham equations}
\author{Gino Biondini$^{1,2}$, Mark A. Hoefer$^3$ and A. Moro$^4$}
\address{$^1$ Department of Mathematics, State University of New York, Buffalo, NY 14260\\
$^2$ Department of Physics, State University of New York, Buffalo, NY 14260\\
$^3$ Department of Applied Mathematics, University of Colorado, Boulder, CO 80303\\
$^4$ Department of Mathematics, Physics and Electrical Engineering, Northumbria University, Newcastle, NE1 8TS, United Kingdom
}
\date{\small\today}

\begin{abstract}
  \bb Reductions of the \eb KP-Whitham system, namely the (2+1)-dimensional hydrodynamic
  system of \bb five \eb equations that describes the slow modulations of
  periodic solutions of the Kadomtsev-Petviashvili (KP) equation, \bb are \eb
  studied.  \bb Specifically, the \eb soliton and harmonic wave limits of the KP-Whitham
  system are considered, \bb which give \eb rise in each case to a four-component
  (2+1)-dimensional hydrodynamic system.  It is shown that a suitable
  change of dependent variables splits the resulting four-component
  systems into two parts: (i) a decoupled, independent two-component
  system comprised of the dispersionless KP equation, (ii) an
  auxiliary, two-component system coupled to the mean flow equations,
  which describes either the evolution of a linear wave or a soliton
  propagating on top of the mean flow.  The integrability of both
  four-component systems is then \bb demonstrated \eb by applying the Haantjes
  tensor test as well as the method of hydrodynamic reductions.
  Various exact reductions of these systems are then presented that
  correspond to concrete physical scenarios.
  \\[1ex]
  \textit{AMS classification codes: 37K40, 74J25, 74J30}
\\
\textit{Keywords: Kadomtsev-Petviashvili, Whitham theory, hydrodynamic systems, integrability.}
\\[1ex]
\today
\end{abstract}

\section{Introduction}
\label{s:intro}

The study of infinite-dimensional integrable systems continues to be
an essential part of contemporary theoretical physics and applied
mathematics, for several reasons.  On the one hand, these systems are
amenable to analytical treatment; they possess a deep mathematical
structure and admit a rich family of exact solutions.  On the other
hand, these systems often arise as model equations in a variety of
concrete physical situations. It is often the case that, even
when the governing equations in a given physical situation are not
integrable, they are close enough to an integrable system that the
properties and solutions of the nearby integrable system provide
useful insight to study more general scenarios.

The majority of known integrable systems are one-dimensional, or, more
precisely, (1+1)-dimensional, meaning systems with one spatial and one
temporal dimension.  An ongoing theme in the last fifty years has been
to extend the theory of integrable systems to multi-dimensional
systems, that is, to systems in more than one spatial dimension.  Of
course multi-dimensional systems are much more challenging than their
one-dimensional counterparts
\cite{AC1991,AS1981,BBEIM1994,Dickey,IR2000,Konopelchenko,NMPZ1984}.
As a distinguished example, consider the Kadomtsev-Petviashvili (KP)
equation \cite{KP1970}, which is arguably the prototypical
(2+1)-dimensional integrable system, and which arises as a governing
equation, for example, in asymptotic approximations of water waves,
plasma physics, cosmology, and condensed matter
\cite{AS1981,IR2000,KP1970}.

While the inverse scattering transform for the Korteweg-de\,Vries
equation (the one-dimensional counterpart of the KP equation)
was originally developed in the 1960s, its development to solve the initial value
problem for a class of functions broad enough to include soliton
solutions required an extensive effort over a period spanning many
decades \cite{IP17p937,JMP44p3309,TMP159p721,TMP165p1237,Dryuma}, and
fundamental questions regarding the dynamics of solutions that are not
simply localized perturbations of soliton solutions are still by and
large unanswered.  The KP equation also possesses a much richer family
of exact solutions, with a more complex mathematical structure
\cite{JMP47p033514,Dickey,hirota1988,JPA2004v37p11169,PNAS2011} and
more complicated physical behavior
\cite{PRL99p064103,JMP47p033514,JPA2004v37p11169,Medina}, than the KdV
equation.

Another ongoing research theme in the last fifty years 
has been the study of dispersionless and semiclassical limits of dispersive evolution equations,
which typically give rise to systems of equations of hydrodynamic type \cite{EH2016}.
Here too, the attempt to extend our knowledge of one-dimensional systems to multi-dimensional ones
has been an ongoing challenge.
For example, with regard to multi-dimensional hydrodynamic systems, 
considerable effort has been devoted over the years to their classification \cite{JPA37p2949,JPA42p035211,JPA39p10803},
the development of effective tests for integrability \cite{ferapontov2004,ferapontov2006},
the search for exact solutions \cite{pla129p223,pla135p167,pavlov2008},
the recent development of a generalized inverse scattering transform for vector fields \cite{manakovsantini_ist},
and the study of the behavior of solutions, 
either numerically \cite{klein} or analytically \cite{manakovsantinibreaking1,manakovsantinibreaking2}.

A useful tool in the study of small dispersion limits is nonlinear
modulation theory, developed by G. B. Whitham in the late 1960's
\cite{Whitham1965,Whitham1976}.  Indeed, Whitham modulation theory has
been applied with enormous success in a wide variety of situations
(e.g., see the review article \cite{EH2016} and the references
therein).  With few exceptions \cite{Bogaevskii,IR2000,Krichever1988},
however, the vast majority of the works that use Whitham's approach
have concerned effectively one-dimensional systems.  A step forward
towards applying Whitham modulation theory to multi-dimensional
systems was recently presented in \cite{PRSA2017}, where Whitham
modulation theory was successfuly generalized to the KP equation, and
a system of equations was derived that governs slow variations of the
periodic solutions of the KP equation.  The authors of \cite{PRSA2017}
referred to these modulation equations as the KP-Whitham system.
Importantly, the validity of the approach and the results of
\cite{PRSA2017} were confirmed by comparing the predictions of the
KP-Whitham system with direct numerical simulation of solutions of the
KP equation, which confirmed that indeed the former correctly captures
the dynamics of the latter in the small dispersion limit.  The results
of \cite{PRSA2017} were then extended to the two-dimensional
Benjamin-Ono equation in \cite{PRE2017} and to other (2+1)-dimensional
equations of KP type in \cite{JPA2018}.
Several important questions were left open in
\cite{JPA2018,PRSA2017,PRE2017}, however, including fundamental issues
about the integrability of the resulting Whitham modulation systems
and their solutions.  The purpose of this work is to present several
new results in this regard.

The rest of this work is organized as follows.  We begin in
section~\ref{s:KP} by reviewing the relation of the one-phase
(genus-1) KP-Whitham system to the solutions of the KP equation as
well as its harmonic wave (elliptic parameter $m \to 0$) and soliton
($m \to 1$) reductions.  We then show that, for both of these special
cases, a suitable change of dependent variables decouples the mean
flow from the rest of the system, thereby transforming these two
special cases of the KP-Whitham system into the dispersionless KP
(dKP) system together with two additional evolution equations that
describe the soliton dynamics (in the case $m\to 1$) or that of a
linear wave packet (in the case of $m\to 0$).  In
section~\ref{s:integrability}, we study the question of the
integrability of the KP-Whitham system by applying both the Haantjes
tensor test \cite{ferapontov2006} as well as the method of
hydrodynamic reductions \cite{ferapontov2004} to the $m\to 0$ and
$m\to 1$ reductions.  We show that the $m\to 0$ and $m \to 1$ cases of
the KP-Whitham system are both completely integrable in the sense of
hydrodynamic reductions.  In section~\ref{s:characteristics}, we show
how the system obtained in the limit $m\to0$ can be completely
integrated by characteristics once a solution of the dKP system is
given.  In section~\ref{s:reductions}, we present several \bb properties
and \eb exact reductions of the system obtained in the limit $m\to1$,
corresponding to different physical scenarios.
We conclude this work in section~\ref{s:remarks} 
with some final remarks.


\section{The KP-Whitham system and its harmonic and soliton limits}
\label{s:KP}

\subsection{The KP equation and the KP-Whitham system}

The KP equation in evolution form is the system
\bse
\label{e:kp}
\begin{align}
&u_t + u u_x + u_{xxx} + \lambda v_y = 0\,,\qquad\qquad\qquad
\label{e:kp1}
\\
&v_x = u_y\,.
\label{e:kp2}
\end{align}
\ese where subscripts $x$, $y$ and $t$ denote partial differentiation,
the dependent variables $u(x,y,t)$ and $v(x,y,t)$ are real-valued, and
$\lambda = \mp1$ identifies respectively the KPI and KPII equations.
Note that the authors of \cite{PRSA2017} used the ``IST-friendly''
normalization in which the nonlinear term in~\eqref{e:kp1} is replaced
with $6uu_x$. They also considered the small-dispersion, hydrodynamic
scaling in which order-one spatial and temporal scales are assumed
slow. This results in a parameter $\epsilon^2$ in front of the
dispersive term $u_{xxx}$ where $0<\epsilon\ll1$ quantifies the
relative strength of dispersive effects compared to nonlinear ones.
Here we use the ``physics-friendly'' normalization without the
coefficient 6 and where order-one length and time scales are
considered fast.  This is equivalent to a scaling of the \bb modulation \eb
variables introduced below.

The exact, elliptic traveling wave solutions of~\eqref{e:kp} are given
in terms of the cnoidal-wave expressions \bse
\label{e:ellipticsoln}
\begin{align}
  &u(x,y,t) = r_1 - r_2 + r_3 + 2(r_2 - r_1)\cn^2 \left[ (\K/\pi)\theta;m \right ]\,,
    \label{e:ugenus1}
  \\
  &v(x,y,t) = qu + p\,, 
\end{align}
\ese
where $r_1,r_2,r_3,q$ and $p$ are constant parameters,
$\cn(\cdot)$ denotes a Jacobian elliptic function, and the rapidly
varying phase $\theta(x,y,t)$ is identified (up to an integration
constant) by
\bse \be
\label{e:thetaderiv}
\theta_x = k\,, \qquad 
\theta_y = l\,, \qquad 
\theta_t = -\omega\,,\qquad\qquad
\ee
with 
\begin{align}
&k = \frac{\pi \sqrt{r_3 - r_1}}{\sqrt{6} \K}\,,\qquad 
q = l/k\,,\qquad
\omega = (V + \lambda q^2)k\,,
\end{align}
with the elliptic parameter $m$ and the velocity parameter $V$
respectively given by
\begin{align}
  \label{eq:1}
&m = \frac{r_2-r_1}{r_3-r_1},\qquad
V = \txtfrac13(r_1 + r_2 + r_3), 
\end{align}
\ese
and $\K = K(m)$ and $\E = E(m)$ denote, respectively, the
complete elliptic integrals of the first {and second}
kind~\cite{nist}.

Note that, in \cite{PRSA2017}, the function $\theta$ was taken to have period 1; here we take it to have period $2\pi$ instead,
which is equivalent to a rescaling of $k$, $l$ and $\omega$.
The reason for the present choice is that, in the harmonic wave limit,
$k$ reduces to the usual wavenumber for trigonometric waves.

It was shown in \cite{PRSA2017}, using a multiple scales expansion that
slow modulations according to \eqref{e:kp} of the above
cnoidal wave solutions are governed by the KP-Whitham system
\bse
\label{e:kpwhitham}
\begin{align}
&\partialderiv{r_j}{t} + (V_j + \lambda q^2)\,\partialderiv{r_j}{x} + 2 \lambda q  \DDy{r_j} + \lambda \nu_j \DDy{q}
  { + \lambda \DDy{p}}
  = 0\,, \qquad j=1,2,3,
\label{e:kpw1}
\\
&\partialderiv{q}{t} + (V_2 + \lambda q^2)\,\partialderiv{q}{x} + 2 \lambda q \DDy{q} 
+ \nu_{4.1} \DDy{r_1} + \nu_{4.3} \DDy{r_3} = 0\,,  
\label{e:kpw2}
\\
\label{e:pcontr}
&{
\partialderiv{p}{x} - (1-\alpha)\DDy{r_1} - \alpha\DDy{r_3} + \nu_5 \partialderiv{q}{x} = 0\,,
}
\end{align}
\ese
where the ``convective'' derivative is defined as 
\begin{equation}
\label{e:DDy}
\frac{D}{D y} = \partialderiv{ }y - q \partialderiv{ }x\,,
\end{equation}
with 
\bse
\label{e:kpwhithamcoeffs}
\begin{align}
V_1 = V - b \frac{\K}{\K - \E}\,,\quad
V_2 = V - b \frac{(1-m) \K}{\E - (1-m) \K}\,,\quad 
V_3 = V + b \frac{(1-m) \K}{m\E}\,,
\label{e:Vdef}
\end{align}
and $V$ as above, as for the KdV equation,
with $b = \frac23(r_2-r_1)$,
and where the remaining coefficients are given by 
\begin{align}
\label{e:nudef}
&\nu_1 = V + \frac{6b}{m} \frac{(1+m)\E-\K}{\K-\E}\,, \qquad
\nu_2 = V + \frac{6b}{m} \frac{(1-m)^2\K - (1-2 m)\E}{\E - (1-m)\K}\,, \\
&\nu_3 = V + \frac{6b}{m} \frac{(2-m)\E -(1-m)\K}{\E}\,, \qquad
\nu_4 = \frac{2m\E}{\E - (1-m) \K}\,, \\
&\nu_{4.1} = \txtfrac16(4-\nu_4)\,, \qquad 
\nu_{4.3} = \txtfrac16(2+\nu_4)\,, \qquad 
 \nu_5 = r_1 - r_2 + r_3\,, \qquad
{\alpha = {\E}/{\K}\,.}
\end{align}
\ese When $p = q = 0$, the resulting system is the (1+1)-dimensional
KdV-Whitham system written in diagonal, Riemann invariant form
\cite{Whitham1965}.  The reduction of~\eqref{e:kpw1}
and~\eqref{e:kpw2} with $p=0$ (but without the corresponding
constraint \eqref{e:pcontr}) was also derived in \cite{grava} using
Lagrangian averaging.

\subsection{Harmonic limit, soliton limit and decoupling of the mean flow}

It was shown in \cite{PRSA2017} that the KP-Whitham
system~\eqref{e:kpwhitham} inherits the invariances of the KP equation
under space-time translations, scaling transformations, Galilean
boosts and pseudo-rotations.  Some of these invariances will be useful
in this work.  It was also shown in \cite{PRSA2017} that the
system~\eqref{e:kpwhitham} admits several distinguished limits and
exact reductions.  In particular, in the ``harmonic limit'', that is,
the limit $r_2 \to r_1$, we have $m \to 0$ in \eqref{eq:1} so that the
cnoidal wave \eqref{e:ugenus1} limits to a vanishing\bb-amplitude \eb
trigonometric or harmonic wave, and the system \eqref{e:kpwhitham}
reduces to \bse
\label{e:kpwhitham1reduction}
\begin{align}
&\partialderiv{r_1}{t} + (2 r_1 - r_3 + \lambda q^2)\,\partialderiv{r_1}{x} + 2 \lambda q  \DDy{r_1} + \lambda r_3 \DDy{q} + \lambda \DDy{p} = 0\,,
\label{e:kpw0a}
\\
&\partialderiv{r_3}{t} + (r_3 + \lambda q^2)\,\partialderiv{r_3}{x} + 2 \lambda q  \DDy{r_3} + \lambda r_3 \DDy{q} + \lambda \DDy{p} = 0\,,
\label{e:kpw0b}
\\
&\partialderiv{q}{t} + (2 r_1 - r_3 + \lambda q^2)\,\partialderiv{q}{x} + 2 \lambda q \DDy{q} 
 + \DDy{r_3} = 0\,, 
\label{e:kpw0c}
\\
&\partialderiv {p}{x} - \DDy{r_3} + r_3 \partialderiv {q}x = 0\,.
\label{e:kpw0d}
\end{align}
\ese Conversely, in the ``soliton limit'', that is, the limit
$r_2\to r_3$, we have $m\to1$ in \eqref{eq:1} so that the cnoidal wave
\eqref{e:ugenus1} limits to a $\mathrm{sech}^2$ profile, and
\eqref{e:kpwhitham} reduces to the system \bse
\label{e:kpwhitham2reduction}
\begin{align}
&\partialderiv{r_1}{t} + (r_1 + \lambda q^2)\,\partialderiv{r_1}{x} + 2 \lambda q  \DDy{r_1} + \lambda r_1 \DDy{q} + \lambda \DDy{p} = 0\,,
\label{e:kpw1a}
\\
&\partialderiv{r_3}{t} + \bigg(\frac13(r_1 + 2 r_3) + \lambda q^2 \bigg)\,\partialderiv{r_3}{x} + 2 \lambda q  \DDy{r_3} + \lambda \frac{4 r_3 - r_1}{3} \DDy{q} + \lambda \DDy{p} = 0\,,
\label{e:kpw1b}
\\
&\partialderiv{q}{t} + \bigg(\frac13(r_1 + 2 r_3) + \lambda q^2 \bigg)\,\partialderiv{q}{x} + 2 \lambda q \DDy{q} 
 + \frac13 \DDy{r_1} + \frac23 \DDy{r_3} = 0\,, 
\label{e:kpw1c}
\\
&\partialderiv {p}{x} - \DDy{r_1} + r_1 \partialderiv {q}x = 0\,.
\label{e:kpw1d}
\end{align}
\ese 
\bb 
Note that both reductions are one-sided limits, and the restrictions 
$r_1 \le r_2 \le r_3$ ($0 \le m \le 1$) are always maintained.
\eb

As we show next, both of these systems describe concrete physical
scenarios.  We first consider the limit $m\to 0$ of the KP-Whitham
system.  Since $r_2 \to r_1$ in this case, the mean flow is simply
given by $\=u = r_3$.  Similarly, $\=v = q\=u + p$.
Rewriting~\eqref{e:kpw0b} and~\eqref{e:kpw0d} in terms of $\=u$ and
$\=v$, we obtain
the two equations
\bse
\label{e:m=0kpwhitham}
\begin{align}
&\u_t + \u \u_x + \lambda \v_y = 0\,,\qquad\qquad\qquad\qquad\quad~
\label{e:kpw3a}
\\
&\v_x = \u_y\,,
\label{e:kpw3b}
\end{align}
which are decoupled from the rest of the system.  
We then note that, in the limit $m\to0$, 
the wavenumber of the cnoidal oscillations in~\eqref{e:ugenus1} 
becomes
$k = 2\sqrt{r_3 - r_1}/\sqrt{6}$.  
Rewriting 
the remaining equations in terms of $\u$, $\v$, $q$ and 
$\k$, we obtain 
\begin{align}
&\k_t + (\u - 3\k^2 - \lambda q^2)\,\k_x + 2\lambda q \k_y + \k\u_x = 0\,,\qquad\quad
\label{e:kpw3c}
\\
&q_t + (\u - 3\k^2 - \lambda q^2)\,q_x  + 2\lambda q q_y + \u_y - q \u_x = 0\,.
\label{e:kpw3d}
\end{align}
\ese
The system of equations \eqref{e:m=0kpwhitham} is
equivalent to the system~\eqref{e:kpwhitham1reduction}.

Note, however, that \eqref{e:kpw3a} and~\eqref{e:kpw3b} are simply the
dispersionless KP (dKP) system.  Therefore, the change of variables
from $(r_1,r_3,q,p)$ to $(\bb \u,\v\eb,\k,q)$ is convenient not only because
it decouples two of the equations from the rest of the system, but
also because it clarifies the physical meaning of the system:
equations \eqref{e:kpw3a} and~\eqref{e:kpw3b} govern the dynamics of
the mean flow, whereas \eqref{e:kpw3c} and~\eqref{e:kpw3d} govern the
dynamics of harmonic waves propagating on this mean flow.  Note also
that the homogeneous system associated of the forced system of
equations \eqref{e:kpw3c} and~\eqref{e:kpw3d} is in diagonal form.
\bb
The decoupling of the mean flow from the remaining modulation variables in
the soliton and harmonic limits is a general property of
(1+1)-dimensional Whitham modulation systems \cite{EH2016}.  Here, we
have shown that this property persists for the (2+1)-dimensional
KP-Whitham modulation system.
\eb

To further elucidate the physical meaning of equations~\eqref{e:kpw3c}
and~\eqref{e:kpw3d}, it is useful to go back to the derivation of the
original KP-Whitham system~\eqref{e:kpwhitham}.  Recall that, when the
KP-Whitham system is used to describe solutions of the KP equation, we
have $\theta_{xy} = \theta_{yx}$.  Hence, \eqref{e:kpw3c}
and~\eqref{e:kpw3d}
are subject to the compatibility condition
\bse
\begin{equation}
  \label{eq:2}
  k_y = (kq)_x \,.\quad~~
\end{equation}
Combining 
\eqref{eq:2} with \eqref{e:kpw3c} and~\eqref{e:kpw3d}, 
we find that the latter two 
equations are equivalent to wave conservation
(that is, 
to the conditions 
$\theta_{xt} = \theta_{tx}$ and $\theta_{yt} = \theta_{ty}$):
\label{eq:consWaves}
\begin{align}
  \label{eq:3}
  &k_t + \omega_x = 0, \\
  \label{eq:4}
  &(kq)_t + \omega_y = 0,
\end{align}
\ese
where $\omega = (u + \lambda q^2) k - k^3$ is the KP linear
dispersion relation.  
Hence, \eqref{e:kpw3c} and~\eqref{e:kpw3d} are simply equivalent to wave conservation
for modulated linear waves.  

A similar result can be obtained in the soliton limit of the
KP-Whitham system, i.e., the limit $r_2 \to r_3$.  In this case, 
$\=u = r_1$, and similar
operations as above 
on \eqref{e:kpw1a} and \eqref{e:kpw1d}
yield
\bse
\label{e:m=1kpwhitham}
\begin{align}
\label{req1}
&\u_t + \u \u_x + \lambda \v_y = 0 \,, \\
\label{req2}
&\u_y - \v_x = 0 \,, \\
\label{req3}
& a_t + \left (\u + \txtfrac13 a - \lambda q^2 \right ) a_x + \lambda\,\big( 2 q a_y + \txtfrac43 a (q_y - q q_x ) \big)  + \txtfrac23 a \u_x = 0 \,,\\
\label{req4}
& q_t + \left (\u + \txtfrac13 a - \lambda q^2 \right ) q_x + 2 \lambda
  q q_y + \txtfrac13  (a_y - q a_x )   + \u_y - q \u_x = 0 \,.
\end{align}
\ese In this case, the variable $a = 2(r_3 - r_1)$ describes the
soliton amplitude, while $q$ defines its inclination in the $xy$-plane
according to the limit $r_2 \to r_3$ of eq.~\eqref{e:ugenus1}
\begin{equation}
  \label{eq:13}
  u(x,y,t) = \u + a \,\mathrm{sech}^2 \left ( \sqrt{\frac{a}{12}} ( x +
    q y - ct) \right ), \quad c = \lim_{k \to 0} \frac{\omega}{k} = \u
  + \frac{1}{3} a + \lambda q^2 .
\end{equation}

Thus, similarly to the \bb harmonic \eb limit, the first two equations, which are
decoupled, describe the dynamics of the mean flow, governed by the dKP
equation, while
\eqref{req3} and \eqref{req4} describe the effect of
the mean field on the soliton's properties.

Note that homogeneous equations associated to equations \eqref{req3} and \eqref{req4} 
\bb(i.e., the reductions of those equations when $\u$ and $\v$ are constant) \eb
are not in diagonal form,
unlike \eqref{e:kpw3c} and \eqref{e:kpw3d}.  On the other hand,
in section~\ref{s:reductions} we will show that it is also possible to write 
\eqref{req3} and \eqref{req4} 
in diagonal form.

The systems~\eqref{e:m=0kpwhitham} and~\eqref{e:m=1kpwhitham} are the
two-dimensional generalizations of the corresponding systems for the
KdV equation that were recently studied theoretically  in
\cite{Congy2018,prl120p144101} and also experimentally  in \cite{prl120p144101}.  
The rest of this work is devoted to the study
of the properties and solutions of the hydrodynamic systems of
equations~\eqref{e:m=0kpwhitham} and \eqref{e:m=1kpwhitham}.


\section{Integrability of the reduced KP-Whitham systems of equations}
\label{s:integrability}

An important open question for the KP-Whitham
system~\eqref{e:kpwhitham} concerns its possible integrability.  On
the one hand, as an exact asymptotic reduction of the KP equation,
which is a completely integrable system, we expect it to also be
completely integrable.  It was stated in \cite{PRSA2017} that the full
KP-Whitham system~\eqref{e:kpwhitham} fails the Haantjes test for
integrability 
\bb
proposed in \cite{ferapontov2006}.  
On the other hand, the formulation of the Haantjies test given in \cite{ferapontov2006} does not directly apply
to KP-Whitham system and its reductions, as we discuss next.
\eb


\subsection{Integrability test via the Haantjes tensor}

In 2006, Ferapontov and Khusnutdinova,  based on the observations of
Gibbons and Tsarev in the context of the \bb Benney moment \eb equations~\cite{gibbonstsarev}, identified a necessary condition for the integrability of a 
(2+1)-dimensional system of hydrodynamic type 
\cite{ferapontov2006}.
%
A necessary condition for the integrability of an $N$-component (2+1)-dimensional quasi-linear system of equations in the form
\be
\@u_T + A(\@u)\,\@u_X + B(\@u)\,\@u_Y = \@0
\label{e:FKsystem} 
\ee
is the integrability of all of its (1+1)-dimensional  hydrodynamic  reductions.
Based on a classical result by Haantjes~\cite{haantjes}, the ($1+1$)-dimensional hyperbolic (with distinct characteristic speeds)  hydrodynamic type system
 
\be
\@u_\tau + M(\@u)\,\@u_\xi = 0
\ee
is diagonalizable if and only if the Haantjes tensor of the matrix $M$ is identically zero.
Recall that the Haantjes tensor $H(\@u)$ of the matrix $M(\@u) = (m^i_j)_{i,j=1,\dots,N}$ is given by
\vspace*{-1ex}
\be
H^i_{jk} = \sum_{p,q=1}^N \big( N^i_{pq} M^p_jM^q_k - N^p_{js} M^i_pM^s_k - N^p_{sk} M^i_pM^s_j + N^p_{jk} M^i_s M^s_p \big)\,,
\ee
where $N(\@u)$ is the Nijenhuis tensor of $M(\@u)$, namely,
\vspace*{-1ex}
\be
N^i_{jk} = \sum_{p=1}^N  \bigg[ M^p_j \partialderiv{M^i_k}{u^p} - M^p_i \partialderiv{M^i_j}{u^p}
  - M^i_p \bigg( \partialderiv{M^p_k}{u^j} - \partialderiv{M^p_j}{u^k} \bigg) \bigg]\,.
\ee
It was then shown in \cite{ferapontov2006} that,
as a result, 
a necessary condition for integrability of the system~\eqref{e:FKsystem} is the vanishing of the Haantjes tensor of the matrix
\be
M = (\lambda I + A)^{-1}(\mu I + B)
\label{e:Mdef}
\ee
for arbitrary values of $\lambda$ and $\mu$, where $I$ is the $N\times N$ identity matrix.

\bb
We emphasize, however, that the Haantjes tensor test cannot be directly applied to the KP-Whitham system~\eqref{e:kpwhitham} or to its reductions
\eqref{e:m=0kpwhitham}
and~\eqref{e:m=1kpwhitham}, because none of 
\eb
these systems are in evolutionary form with respect to the variable $t$.
This is because no temporal derivative is present in
neither~\eqref{e:kpw3b} nor~\eqref{req2}.  
In other words, introducing
$\@u = (\u,a,q,\v)^T$, \eqref{e:m=0kpwhitham} and~\eqref{e:m=1kpwhitham}
yield \bb systems \eb of the form 
\be 
\~I_4\@u_t + \~A(\@u)\,\@u_x +
\~B(\@u)\,\@u_y = \@0\,,
\label{e:kpsystem} 
\ee 
where $\~I_4 = \diag(1,1,1,0)$.  Thus, we cannot simply identify
$(X,Y,T)$ in \eqref{e:FKsystem} with $(x,y,t)$.  
\bb
A~similar issue arises with the full KP-Whitham system~\eqref{e:kpwhitham}. 
Accordingly, the question of whether the full KP-Whitham system is integrable is still open. 
Such a question is outside the scope of this work, and we therefore leave it for future study.
\eb

\bb 
Instead, in this work we limit ourselves to the study of the four-component 
reductions~\eqref{e:m=0kpwhitham} and~\eqref{e:m=1kpwhitham}.
One can obviate the above problems for the four-component reductions~\eqref{e:m=0kpwhitham} and~\eqref{e:m=1kpwhitham}
by writing each of them 
\eb
as an
evolutionary system with respect to either of the variables $x$ or~$y$.  
In the first case, we identify $(X,Y,T)$ \bb in~\eqref{e:FKsystem} \eb with $(t,y,x)$ \bb in~\eqref{e:kpsystem}\eb,
respectively, obtaining $A = \~A^{-1}\~I_4$ and $B = \~A^{-1}\~B$.  In
the second case, we identify $(X,Y,T)$ with $(t,x,y)$, respectively,
obtaining $A = \~B^{-1}\~I_4$ and $B = \~B^{-1}\~A$.  We find that the
Haantjes tensor of the matrix $M$ associated to either of these
systems via~\eqref{e:Mdef} is identically zero, which suggests that
these systems might be completely integrable.
(The relevant
  calculations, which were performed with Mathematica, are relatively
  straightforward but rather tedious, and for this reason are omitted
  here.)

On the other hand, the Haantjes test is only a necessary condition for
integrability.  Next, we discuss a more conclusive test.

\subsection{Integrability test via hydrodynamic reductions}

A (2+1)-dimensional quasilinear system is said to be ``integrable'' if
it admits infinitely many reductions into a pair of compatible
$N$-component one-dimensional systems in Riemann invariants
\cite{ferapontov2004}.  Exact solutions described by these reductions,
known as nonlinear interactions of planar simple waves, can be viewed
as a natural dispersionless analogue of $N$-gap,
quasi-periodic solutions. 

 In this section, we focus on the case corresponding to the soliton
limit ($m \to 1$) of the KP-Whitham system, namely the
system~\eqref{e:m=1kpwhitham}, in the case $\lambda=1$ (i.e., for the
KPII equation). The harmonic limit ($m \to 0$) is considered
separately in the next section and, since the mean flow equations are
integrable and the remaining equations can be solved in full
generality, no integrability test is required.  We look for
$N$-component reductions, i.e., solutions in the form of arbitrary
functions of $N$ variables $R^1,\dots,R^N$: \bse
\begin{align}
  \label{eq:5}
  &\u = \bb\u\eb(R^{1},\dots,R^N)\,,\qquad 
    \v = \bb\v\eb(R^{1},\dots,R^N)\,,
  \\
  \label{eq:6}
  &a = a(R^{1},\dots,R^N)\,,\qquad 
    q = q(R^{1},\dots,R^N)\,, 
\end{align}
\ese
where
\bse
\begin{align}
&R^i_{y} = \lambda^i(R^{1},\dots,R^N) R^i_{x}\,,\\
&R^i_{t} = \mu^i(R^{1},\dots,R^N) R^i_{x}\,
\end{align}
\ese (with no sum on repeated indices), \bb subject to \eb the compatibility
conditions
\begin{align}
\frac{\lambda^i_{j}}{\lambda^i - \lambda^{j}}  =\frac{\mu^i_{j}}{\mu^i - \mu^{j}}\,,
\end{align}
where for brevity we used the notation $\lambda^i_{j} = \partial \lambda^i / \partial R^{j}$ 
and similarly for $\mu_j^i$ as well as $\bb\u_j\eb$ and $\bb\v_j\eb$.
Substituting into \eqref{req1} and~\eqref{req2}, we obtain
\begin{equation}
\label{dKPdisp}
\bb\v_i\eb = \lambda^i \bb\u_i\eb\,,\qquad 
\mu^i = - \bb\u\eb - ( \lambda^i )^2\,.
\end{equation}
The compatibility condition of the system~\eqref{dKPdisp} leads to the Gibbons-Tsarev system for the dKP equation \cite{gibbonstsarev},
namely, 
\bse
\begin{align}
\bb\u_{ij}\eb &= \frac{2 \bb\u_i \u_{j}\eb}{(\lambda^{j} - \lambda^i)^2}\,, 
\\
\lambda^i_{j} &= \frac{\bb\u_{j}\eb}{\lambda^{j} - \lambda^i}\,,
\end{align}
\ese 
which is automatically in involution.  
Now using the assumptions
\eqref{eq:5} and \eqref{eq:6} in~\eqref{req3} and~\eqref{req4}, we
obtain 
\bse
\label{e:aqreductions}
\begin{align}
a_i &= - \frac{2 a \bb\u_i\eb}{a - q^2 + 2 q \lambda^i - (\lambda^i)^2}\,,
\\
q_i &=  - \frac{(\lambda_i - q)\bb\u_i\eb}{a- q^2 + 2 q \lambda^i - (\lambda^i)^2}\,.
\end{align}
\ese 
We verify by direct calculation that the compatibility conditions
\be 
a_{ij} = a_{ji}\,,\qquad q_{ij} = q_{ji} 
\ee 
are identically satisfied modulo all equations above.
(As before, the relevant
  calculations were performed with Mathematica; they are relatively
  straightforward but rather tedious, and are omitted here for
  brevity.)

Therefore, the system~\eqref{e:m=1kpwhitham} is integrable in the sense
of hydrodynamic reductions.  Moreover, the
system~\eqref{e:aqreductions} of ordinary differential equations
(ODEs) allows us to obtain the solutions for $a$ and $q$ associated to
any $N$-component reduction of the dKP equation.
%


\section{General solution  of the harmonic wave limit of the KP-Whitham system}
\label{s:characteristics}

 We could perform similar calculations as in
section~\ref{s:integrability} for the harmonic limit ($m\to0$) of the
KP-Whitham system.  In this case, however, we next prove that the
system is integrable by integrating it exactly.  For convenience, let
us rewrite  the Whitham system in the  harmonic wave 
limit~\eqref{e:m=0kpwhitham}:
\begin{subequations}
\label{e:harmonicsystem}
\begin{align}
\label{dKPi}
&\bb\u\eb_t + \bb\u\eb \bb\u\eb_x + \lambda \bb\v\eb_y = 0 \\
\label{dKPii}
&\bb\v\eb_x = \bb\u\eb_y \\
&k_t + (\bb\u\eb -  3k^2
 -  \lambda q^2)\,k_x + 2\lambda q k_y + k \bb\u\eb_x = 0 
\label{e:hs3}
\\
&q_t + (\bb\u\eb -  3k^2 -  \lambda q^2)\,q_x  + 2\lambda q q_y +  \bb\u\eb_y - q \bb\u\eb_x  = 0\,.
\label{e:hs4}
\end{align}
\end{subequations}
Introducing the characteristic curves defined via the system
\begin{equation}
\label{characteristics}
\deriv{x}{t} = \bb\u\eb -  3k^2  -  \lambda q^2  \,,\qquad 
\deriv{y}{t} = 2 \lambda q\,,
\end{equation}
we have that the functions $a(x(t),y(t),t)$ and $q(x(t),y(t),t)$
evaluated along the characteristics satisfy the ODEs
\begin{equation}
\label{fderaq}
\deriv{k}{t} = - k \bb\u\eb_{x} \,,\qquad 
\deriv{q}{t} = - \bb\u\eb_{y} + q \bb\u\eb_{x}\,.
\end{equation}
Hence, given any solution $(\bb\u\eb,\bb\v\eb)$ to the integrable dKP
equation~(\ref{dKPi}-\ref{dKPii}), the general solution is obtained by
direct integration of the equations~(\ref{fderaq}) along the
characteristics.  There is a large class of known reductions and
explicit solutions of the dKP system
\cite{pla129p223,pla135p167,KamPav,KMM}, which could be used to obtain
the corresponding solutions of the associated
system~\eqref{characteristics}.

As an illustrative example, let us consider a ``background'' represented by the following solution to the dKP equation
\begin{equation}
\bb\u\eb = \alpha y\,, \qquad 
\bb\v\eb = \alpha x\,.
\end{equation}
where $\alpha$ is a constant. 

\bb In this case, \eb the characteristic system given by 
equations~(\ref{characteristics}) and~(\ref{fderaq}) can be integrated
explicitly, providing the solution in parametric form \bse
\begin{align}
& x(t)  = - \txtfrac{2}{3} \lambda \alpha^{2} t^{3} + 2 \alpha \lambda q_{0} t^{2} + (\alpha y_{0} - \lambda q_{0}^{2} - 3k_{0}^2) t + x_{0}\,,
\label{e:charsoln1}
\\
& y(t)  = - \lambda \alpha t^{2} + 2 \lambda q_{0} t + y_{0}\,,
\label{e:charsoln2}
\\
& \bb k(x(t),y(t),t)  = f( x_0,y_0) \,, \eb
\label{e:charsoln3}
\\
& q(x(t),y(t),t)  = g( x_0,y_0) - \alpha t\,,
\label{e:charsoln4}
\end{align}
\ese
with initial conditions  $x(0) = x_0$, $y(0) = y_0$ and
\[
k(x_0, y_0,0) = f(x_{0},y_{0})\,,\quad 
q(x_0,y_0,0) = g(x_{0},y_{0})\,,
\]
and where, for brevity in the above formulae, we use the shorthand
notation
\[
k_{0} = k_{0}(x_0,y_0)\,,\qquad 
q_{0} = q_{0}(x_0,y_0)\,.
\]

It is important to note that initial conditions must be taken in the class specified by the compatibility condition~(\ref{eq:2}) and therefore $f_{y} = (g f)_{x}$.

If desired, we could now solve \eqref{e:charsoln1} and \eqref{e:charsoln2} to obtain $x_0$ and $y_0$ explicitly as functions of $(x,y,t)$ and then substitute
the resulting expressions in \eqref{e:charsoln3} and \eqref{e:charsoln4} to obtain explicit expressions for $k(x,y,t)$ and $q(x,y,t)$.

It is important to realize that the key observation that allows us to find the general solution of the system~\eqref{e:harmonicsystem}
are that: 
(i)~\eqref{dKPi} and~\eqref{dKPii} are decoupled from \eqref{e:hs3} and~\eqref{e:hs4}, and 
(ii)~the characteristic speeds in~\eqref{e:hs3} and those in~\eqref{e:hs4} coincide.
The first of these properties is also common to the KP-Whitham system in the soliton limit~\eqref{e:m=1kpwhitham}.
The second, however, is not, and this precludes the possibility of solving~\eqref{e:m=1kpwhitham} in the same way.


\section{ Diagonalization and exact reductions of the soliton limit of the KP-Whitham system}
\label{s:reductions}

As discussed above, it is not possible to find the general solution of
the system in the soliton limit as in section~\ref{s:characteristics}.
Nonetheless the $m\to 1$ system~\eqref{e:m=1kpwhitham} admits various
exact reductions that describe interesting physical scenarios.  After
studying the characteristic speeds, we discuss these reductions.

\subsection{Hyperbolicity and characteristic speeds}
\label{sec:hyperb-char-speeds}

The complete modulation system in the soliton limit is \bb given by \eb
\eqref{req1}--\eqref{req4}.  
\bb 
As we mentioned before, this system, 
as well as
the subsystem given by the dKP system \eqref{req1} and \eqref{req2}, is 
\eb 
not evolutionary with respect to the physical time $t$
\cite{briohunter1992,keyfitz}, but they can be written in evolutionary
form, for instance with respect to the variable $y$.  Because dKP
decouples from the soliton's evolution, it will be useful to consider
the soliton evolution equations \eqref{req3}, \eqref{req4} as an
inhomogeneous quasi-linear system for $\@u = (a,q)^T$ \bse
\label{e:forcedsoliton}
\begin{equation}
  \label{eq:11}
  \@u_t + A(\@u) \@u_x + B(\@u) \@u_y +
  \@b = 0,
\end{equation}
with
\begin{equation}
  \label{eq:12}
  A(\@u) =
  \begin{pmatrix}
    \bb\v\eb+\frac{1}{3} a - \lambda q^2 & -\frac{4}{3} \lambda a q \\
    -\frac{1}{3} q & \bb\v\eb+\frac{1}{3} a - \lambda q^2
  \end{pmatrix}, \quad
  B(\@u) =
  \begin{pmatrix}
    2 \lambda q & \frac{4}{3} \lambda a  \\
    \frac{1}{3} & 2 \lambda q
  \end{pmatrix},
\end{equation}
\ese
and inhomogeneity $\@b = (\frac{2}{3} a \bb\v\eb_x,\bb\v\eb_y-q \bb\v\eb_x)^T$
due to the independent mean flow $\bb\u\eb$ satisfying dKP \eqref{req1},
\eqref{req2}.  The variable $t$ is a timelike variable for the
homogeneous part of \eqref{eq:11} and we now analyze the
characteristic speeds of this system.

We study plane wave solutions $\@v\, e^{i(\bb K\eb x + \bb\Lambda\eb y - \bb\Omega\eb t)}$ to~\eqref{eq:11} linearized about the constant $\@u$ and frozen mean
flow $\bb\u\eb$, which yields the generalized eigenvalue problem
\begin{equation}
  \label{eq:9}
  \left ( \bb K \eb A(\@u) + \bb \Lambda \eb B(\@u) - \bb \Omega \eb I_2
  \right ) \@v = 0 ,
\end{equation}
where $I_2$ is the $2\times 2$ identity matrix.  The roots
of the characteristic polynomial are
\begin{equation}
  \label{eq:10}
  \bb\Omega\eb_\pm = \bb K\eb \left (\bb\u\eb + \frac{1}{3}a - \lambda q^2 \right ) + 2 \bb\Lambda\eb
  \lambda q \pm \frac{2}{3} \left |\bb\Lambda\eb - \bb K\eb q \right |\sqrt{\lambda a} 
\end{equation}
and the associated eigenvectors are
\begin{equation}
  \label{eq:8}
  \@v_\pm =
  \begin{pmatrix}
    \pm 2 \left | \bb\Lambda\eb - \bb K\eb q \right | \sqrt{\lambda a} \\
    \Lambda - K q
  \end{pmatrix}
\end{equation}
Therefore, the homogeneous part of the modulation system \eqref{eq:11}
is strictly hyperbolic if and only if $\lambda > 0$ and $\bb\Lambda\eb \ne \bb K\eb q$.
When $\lambda < 0$, the modulation system is elliptic, which is the
modulation theory manifestation of the well-known fact that line
soliton solutions to KPI exhibit a transverse instability
\cite{KP1970}.  From now onward, we will only consider the hyperbolic case
$\lambda = 1$.

We now turn our attention to the characteristic speeds.  Fixing the
propagation direction $\@n = (\cos \theta,\sin \theta)^T$, the
characteristic speeds $\lambda_\pm$ are determined from \eqref{eq:10}
with the identification $\bb K\eb = \cos \theta$, $\bb\Lambda\eb = \sin \theta$, and
$\lambda_\pm = \Omega_\pm$ so that
\begin{equation}
  \label{eq:14}
  \lambda_\pm =  \left ( \bb\u\eb + \frac{1}{3} a - q^2 \right ) \cos \theta
  + 2 q \sin \theta \pm \frac{2}{3} \sqrt{a} \left | \sin \theta - q
    \cos \theta \right | .
\end{equation}
We observe that $\lambda_+ = \lambda_-$ if and only if
$q = \tan \theta$, i.e., the soliton propagation direction coincides
with the characteristic direction.  Thus, the homogeneous part of
\eqref{eq:11} is strictly hyperbolic if and only if it has real
characteristic speeds ($\lambda > 0$) and $q \ne \tan \theta$.  We
compute
\begin{equation}
  \label{eq:15}
  \nabla_{(a,q)} \lambda_\pm \cdot \@v_\pm = \frac{8}{3}\left ( \sin
    \theta - q \cos \theta \right ),
\end{equation}
so that the soliton modulation system \eqref{eq:11} is both strictly
hyperbolic and genuinely nonlinear if and only if $\lambda > 0$ and
$q \ne \tan \theta$.

\subsection{Diagonalization of the homogeneous soliton KP-Whitham system}
\label{sec:bb-diag-homog}

As mentioned earlier, the homogeneous system associated to the
KP-Whitham subsystem \eqref{req3} and \eqref{req4} in the soliton
limit is not in diagonal form, unlike that of the KP-Whitham subsystem
\eqref{e:kpw3c} and \eqref{e:kpw3d} in the harmonic limit.  On the
other hand, this structural discrepancy can be resolved by performing
the change of dependent variables $(a,q)\mapsto(w_1,w_2)$, with
\begin{equation}
\label{wvar}
a = \txtfrac14(w_1 - w_2)^2\,,\qquad 
q = \txtfrac12(w_1 + w_2)\,,
\end{equation}
 which is inverted by
\begin{equation}
  \label{eq:7}
w_1 = q - \sqrt{a}\,,\qquad w_2 = q + \sqrt{a}\,.
\end{equation}
This way, 
\eqref{req3} and~\eqref{req4} take the following diagonal
form
\begin{equation}
  \partialderiv{w_j}t + \@v_j\cdot\,\bfnabla w_j  +  \bb\u\eb_y - b_j
  \,\bb\u\eb_x  = 0\,,\qquad j =1,2\,,
\label{e:m=1diagonal}
\end{equation}
where 
\vspace*{-1ex}
\begin{align}
&\bfnabla = \bigg(\partialderiv{}x,\partialderiv{}y\bigg)\,,
\quad
\@v_1 = \big( \bb\u\eb - w_1 b_2, 2 b_1 \big)^T\,, \quad
  \@v_2 = \big( \bb\u\eb -  w_2 b_1,2 b_2 \big)^T\,,
  \nonumber \\
&
b_1 = \txtfrac13 (2 w_1 + w_2)\,, \quad b_2 = \txtfrac13 (w_1 + 2 w_2) .
\label{e:m=1diagcoeffs}
\end{align}
In these new variables, we identify the characteristic velocities
$\lambda_{1,2} = \@v_{1,2} \cdot (\cos \theta,\sin \theta)$.

We remark that the matrices $A$ and $B$ in \eqref{eq:11},
\eqref{eq:12} have the same eigenvectors and commute.
\bb This is \eb a property of
multidimensional hyperbolic systems that has been leveraged to obtain
uniqueness \cite{fridlefloch2006} and stability \cite{dafermos1995}
results.  Despite having been studied mathematically, this is
apparently the first example of a physically derived hyperbolic system
in multiple dimensions in which the coefficient matrices exhibit this
property \cite{fridlefloch2006}.


\subsection{Constant mean flow}
\label{sec:constant-mean-flow}

Next we consider the $m\to 1$ system~\eqref{e:m=1kpwhitham} in the
case when the mean flow is constant, i.e., $\bb\u\eb=\const$ and
$\bb\v\eb=\const$.  Without loss of generality, we can then set $\bb\u\eb=\bb\v\eb=0$
thanks to the invariances of the KP equation and of the KP-Whitham
system.  The system then reduces to the two-component
(2+1)-dimensional hydrodynamic system of equations
\eqref{e:forcedsoliton} with zero inhomogeneity $\@b = 0$
\bse
\label{e:2component}
\begin{equation}
  \label{eq:19}
  \@u_t + A(\@u) \@u_x + B(\@u) \@u_y  = 0,
\end{equation}
with
\begin{equation}
  \label{eq:20}
  A(\@u) =
  \begin{pmatrix}
    \frac{1}{3} a - q^2 & -\frac{4}{3}  a q \\
    -\frac{1}{3} q & \frac{1}{3} a -  q^2
  \end{pmatrix}, \quad
  B(\@u) =
  \begin{pmatrix}
    2  q & \frac{4}{3}  a  \\
    \frac{1}{3} & 2  q
  \end{pmatrix} .
\end{equation}
\ese This system belongs to the class studied in~\cite{JPA37p2949} and can be written in the diagonal form
(cf.~eqs.~\eqref{e:m=1diagonal})
\bse
\begin{equation}
  \partialderiv{w_j}t + \@V_j\cdot\,\bfnabla w_j = 0\,,\qquad j =1,2\,,
  \label{e:m=1diagonalhomogeneous}  
\end{equation}
where $w_1 = q - \sqrt{a}$ and $w_2 = q + \sqrt{a}$ as in~\eqref{eq:7} and
\vspace*{-1ex}
\begin{align}
&\bfnabla = \bigg(\partialderiv{}x,\partialderiv{}y\bigg)\,,
\quad
\@V_1 = \big( - w_1 b_2, 2 b_1 \big)^T\,, \quad
  \@V_2 = \big( -  w_2 b_1,2 b_2 \big)^T\,,
  \nonumber \\
&
b_1 = \txtfrac13 (2 w_1 + w_2)\,, \quad b_2 = \txtfrac13 (w_1 + 2 w_2) .
\label{e:m=1diagcoeffshomogeneous}
\end{align}
\ese
This system inherits all of the properties (hyperbolicity, etc)
discussed in section \ref{sec:hyperb-char-speeds} for the homogeneous
part of~\eqref{e:forcedsoliton}.

\subsection{One-dimensional fields}

Let us consider the modulation equations~\eqref{e:m=1kpwhitham} in the
case $m\to 1$ when no $y$ dependence is present in the problem.  In
this case, the dependent variable $\v$ is constant and drops out of the
system.  Equations \eqref{e:m=1kpwhitham} then become \bse
\label{e:yindependent}
\begin{equation}
\@u_t + A\,\@u_x = 0\,,
\end{equation}
where 
\begin{equation}
\@u(x,t) = (\bb\u\eb,a,q)^T,\qquad
A = \begin{pmatrix}
  \bb\u\eb & 0 & 0 \\
  \frac{2}{3} a & \bb\u\eb + \frac{a}{3} - q^2 & - \frac{4}{3}
  a q 
  \\
  -q & -\frac{1}{3} q & \bb\u\eb + \frac{a}{3} - q^2
  \end{pmatrix}\,.
\label{e:Acoeff}
\end{equation}
\ese It is important to realize that, even though the spatial
dependence in~\eqref{e:yindependent} is one-dimensional, the geometry
of the resulting solutions for the KP-equation is still
two-dimensional.  This is because the variable $q$ describes the slope
of the soliton in the $xy$-plane.  Thus, solutions for which $q$ is
not constant describe changes in the soliton slope as a function of
$x$.

In the previous two sections, we analyzed the quasi-linear soliton
modulation equations in isolation from dKP evolution because dKP is a
nonlocal equation.  The full modulation system is degenerate as a
hyperbolic system in time.  Here, we can consider the full
soliton-mean flow modulation system because the dKP equation reduces
to the inviscid Burgers equation.  The benefits of considering the
full soliton-mean flow modulation system, despite Burgers equation
being decoupled, have been recently identified \cite{prl120p144101}.

Since the original system~\eqref{e:m=1kpwhitham} is completely
integrable, the reduced system~\eqref{e:yindependent} is too, and can
therefore be diagonalized.  The eigenvalues of the coefficient matrix
in \eqref{e:Acoeff} are given by
$\mathbf\Lambda = (\lambda_{\bb\u\eb},\lambda_+,\lambda_-)$, with
\begin{equation}
  \lambda_{\bb\u\eb} = \bb\u\eb, \quad 
  \lambda_+ = \bb\u\eb + \frac{a}{3} - q^2 - \frac{2}{3} \sqrt{a q^2}, \quad 
  \lambda_- = \bb\u\eb + \frac{a}{3} - q^2 + \frac{2}{3} \sqrt{ a q^2} .
\label{eq:31}
\end{equation}

In what follows, we assume that $q \ge 0$.  
%
%
The ordering of the characteristic velocities is determined by the relative magnitudes of $a$ and $q$.  
In particular, the following special cases arise:
\begin{enumerate}
\item 
$q^2 = a/9$, implying $\lambda_{\bb\u\eb} = \lambda_+$,
\item
$q = 0$, implying $\lambda_+ = \lambda_-$,
\item
$q^2 = a$, implying $\lambda_{\bb\u\eb} = \lambda_-$.
\end{enumerate}
Thus, the system~\eqref{e:yindependent} is non-strictly hyperbolic.
The left eigenvectors associated to each eigenvalue of $A$ are
\begin{equation}
  \label{eq:33}
  \@v_{\bb\u\eb} =
  \begin{pmatrix}
    1 \\ 0 \\ 0
  \end{pmatrix}, \quad 
  \@v_+ =
  \begin{pmatrix}
    \frac{2\sqrt{a}}{\sqrt{a}+q} \\ 1 \\ 2 \sqrt{a}
  \end{pmatrix}, \quad 
  \@v_- =
  \begin{pmatrix}
    \frac{2\sqrt{a}}{\sqrt{a}-q} \\ 1 \\ -2 \sqrt{a}
  \end{pmatrix} .
\end{equation}
We therefore recognize $R_{\bb\u\eb} = \bb\u\eb$ as one Riemann invariant for the system~\eqref{e:yindependent}.  
Taking the dot product of~\eqref{e:yindependent}
with $\@v_+$ yields the characteristic form
\begin{equation}
  \label{eq:35}
  \mathrm{d}\bb\u\eb + \frac{1}{2} \mathrm{d}\left ( q + \sqrt{a}
  \right )^2 = 0 ,
\end{equation}
which can be integrated to give another Riemann invariant
\begin{equation}
  \label{eq:36}
  R_+ = \bb\u\eb + \frac{1}{2}(q + \sqrt{a} )^2 .
\end{equation}
The dot product of \eqref{e:yindependent} with $\@v_-$ gives the third Riemann invariant
\begin{equation}
  \label{eq:37}
  R_- = \bb\u\eb + \frac{1}{2}(q - \sqrt{a} )^2 .
\end{equation}
The Riemann invariants $R_\pm$ for this soliton mean-flow modulation
system generalize the diagonalizing transformation \eqref{eq:7}
according to $R_\pm = \bb\u\eb + \frac{1}{2} w_{1,2}^2$ that was utilized
for the isolated soliton modulation system \eqref{eq:11}.

Note that we have the ordering 
\begin{equation}
  \label{eq:43}
  R_{\bb\u\eb} \le R_- \le R_+ .
\end{equation}
Moreover, \eqref{eq:43} implies the following degenerate cases:
\begin{enumerate}
\item $\sqrt{a} = \pm q$ if and only if $R_u = R_\mp$,
\item $q = 0$ if and only if $R_+ = R_-$.
\end{enumerate}

Summarizing, and denoting $(R_1,R_2,R_3) = (R_{\bb\u\eb},R_+,R_-)$,
the system \eqref{e:yindependent} can be written in Riemann invariant form as
\begin{equation}
\label{e:38}
\partialderiv{R_j}t + \lambda_j \partialderiv{R_j}x = 0 , 
\qquad j = 1,2,3,
\end{equation}
where
\bse
\begin{align}
\label{e:39}
\lambda_1 &= R_1, 
\\
\label{e:40}
\lambda_2 &= \frac{5}{3} R_1 - \frac{2}{3}\left ( R_2 + 
        2\sigma\sqrt{(R_2-R_1)(R_3-R_1)} \right ), 
\\
\label{e:41}
\lambda_3 &= \frac{5}{3} R_1 - \frac{2}{3}\left ( R_3 + 
        2\sigma\sqrt{(R_2-R_1)(R_3-R_1)} \right ) ,
\end{align}
\ese
with $\sigma = \mathrm{sgn}\,(q^2-a)$.  
Note that the mapping from the Riemann invariants $R_1,R_2,R_3$ 
to the modulation parameters $u$, $a$, and $q$ is multivalued:
\bse
\begin{align}
  \label{eq:46}
  u &= R_1, \\
  \label{eq:47}
  a &= -R_1 + \frac{1}{2} \left ( R_2 + R_3 \pm 2 \sqrt{(R_2
      -R_1)(R_3-R_1)} \right ), \\
  \label{eq:48}
  q^2 &= -R_1 + \frac{1}{2} \left ( R_2 + R_3 \mp 2 \sqrt{(R_2
        -R_1)(R_3-R_1)} \right ) .
\end{align}
\ese The branching occurs precisely when $a = q^2$ or $R_1 = R_3$.
Note also how, once more, the dynamics of the mean flow (given by
$\bb\u\eb=R_1$) is decoupled from the rest of the system.

In the regime of strict hyperbolicity, the
system \eqref{e:yindependent} is genuinely nonlinear because
$\partial \lambda_j/\partial R_j \ne 0$, $j = 1,2, 3$ if and only if
$q^2 \notin \{0,\frac{1}{9}a,a\}$.

Finally, we note that the system~\eqref{e:m=1kpwhitham} also admits a
self-consistent reduction when all fields are independent of $x$.  In
this case, both $\bb\u\eb$ and $\v$ drop out of the system, which then reduces
to the one-dimensional 2-component system for $a$ and $q$
\begin{equation}
  \label{eq:16}
  \begin{split}
    a_t + 2 q a_y + \frac{4}{3} a q_y &=0, \\
    q_t + 2 q q_y + \frac{1}{3} a_y &= 0 ,
  \end{split}
\end{equation}
which is equivalent to the diagonalizable (1+1)-dimensional isentropic
dynamics of a monoatomic gas with the ratio of specific heats
$\gamma = 5/3$ after the velocity and density transformation
$U = 2 q$, $\rho = \frac{4}{27} a^{3/2}$, respectively
\begin{equation}
  \label{eq:17}
  \begin{split}
    \rho_t + (\rho U)_y &= 0, \\
    U_t + UU_y + \frac{1}{\rho} P(\rho)_y &= 0,
  \end{split}
\end{equation}
with pressure law $P(\rho) = \frac{3}{5} \rho^{5/3}$.  
\bb
The modulation
system \eqref{eq:16} had been previously obtained in \cite{neu} by an
alternative approach using soliton perturbation theory.
\eb

\section{Final remarks}
\label{s:remarks}

In conclusion, we studied the $m\to 0$ and $m \to 1$  limits  of
the KP-Whitham system of equations.  We showed that both of  the
resulting  systems are completely integrable, and we discussed some
of their exact reductions and some exact solutions.  Both of the
systems~\eqref{e:m=0kpwhitham} and~\eqref{e:m=1kpwhitham} are novel to
the best of our knowledge.  Indeed, the one-dimensional
reduction~\eqref{e:yindependent} and the two-component
reduction~\eqref{e:2component} are also novel to the best of our knowledge.

 Recall that the first half of the systems~\eqref{e:m=0kpwhitham}
and~\eqref{e:m=1kpwhitham} is comprised of the dKP equation.  In
section~\ref{s:characteristics}, we showed that the solution of the
systems~\eqref{e:m=0kpwhitham} can be obtained once $\u$ and $\v$ are
given.  Importantly, however, it does not appear to be possible  in
general to find solutions of the system~\eqref{e:m=1kpwhitham} by
expressing $a$ and~$q$ simply in terms of $\u$ and~$\v$.  This
suggests that the system of equations~\eqref{e:m=1kpwhitham}  is
 not simply a trivial extension of the dKP equation, but  is
instead a novel integrable system in its  own right.  Indeed, 
as we showed in section~\ref{s:reductions},  even  in the case
 when $\u=\v=0$, it is necessary to solve a nontrivial 2-component
(2+1)-dimensional system of hydrodynamic equations to obtain $a$ and
$q$.

The results of this work open up a number of interesting questions.
What is the Lax pair of the (2+1)-dimensional hydrodynamic
systems~\eqref{e:m=0kpwhitham}, \eqref{e:m=1kpwhitham} and
\eqref{e:2component}?  A related question is whether one could use the
recent generalization of the inverse scattering transform for vector
fields  developed by Manakov and Santini
\cite{manakovsantini_ist,manakovsantinibreaking1,manakovsantinibreaking2}
to solve the initial value problem for these hydrodynamic systems.
Another interesting question is whether there exist further exact
reductions of the systems~\eqref{e:m=0kpwhitham}
and~\eqref{e:m=1kpwhitham} and whether further exact solutions can be
obtained, perhaps using similar methods as in \cite{pavlov2008}.
Finally, an important unresolved question concerns the possible
integrability of the full KP-Whitham system~\eqref{e:kpwhitham}.

From a more practical point of view, an important avenue for further
work is the use of the KP-Whitham system to study physically
interesting scenarios.  Even without using the full system, the
reductions studied in this work appear to be very promising in this
respect.  For example, the soliton limit~\eqref{e:m=1kpwhitham} can
allow researchers to study the evolution of initial conditions for the
KP equation that are not purely solitonic in nature, a problem that
appears to still be beyond the capabilities of the state-of-the-art
inverse scattering transform, and which, apart from some numerical
attempts \cite{sapm123p375,kaokodama}, has therefore remained
completely open so far.  Indeed, note that even the one-dimensional
reduction~\eqref{e:yindependent} can be very useful in this regard.
For example, piecewise constant initial conditions in
\eqref{e:yindependent} can be used to study the evolution of scenarios
in which an initial line soliton with a certain amplitude and slope,
located in the first quadrant of the $xy$-plane, is connected at the
origin to a soliton with a different amplitude and slope, located in
the third quadrant of the $xy$-plane.  More complicated scenarios that
cannot be represented with the one-dimensional reduction can then be
studied using the two-dimensional reduction~\eqref{e:2component}.
%


\section*{Acknowledgments}

We thank Eugeny Ferapontov and Michelle Maiden for many interesting
discussions on related topics.  This work was partially supported by
the National Science Foundation under grant numbers DMS-1615524,
DMS-1614623 (G.B.) and DMS-1517291 (M.A.H.), by the Royal Society under the International Exchanges scheme grant number IES\textbackslash R2\textbackslash 170116 (G.B. and A.M.) and by the Leverhulme
Trust Research Project Grant RPG-2017-228 (A.M.)


\def\reftitle#1{``#1''}
\def\journal#1#2{\begingroup\frenchspacing\textit{#1}\/\unskip\, \textbf{\ignorespaces #2}\endgroup}
\def\href#1{\relax}
\let\title=\reftitle

\end{document}